\documentclass[prb,aps,twocolumn,epsfig,eqsecnum,showpacs]{revtex4}
\usepackage{graphicx,epsf}
\usepackage{wasysym}
\usepackage{bm}
\usepackage{pstricks}
\usepackage{psfrag}

\newcommand{\ashtray}{flippable supertetrahedron}
\newcommand{\ashtrays}{flippable supertetrahedra}
\newcommand{\superashtray}{flippable (super)$^2$tetrahedron}
\newcommand{\superashtrays}{flippable (super)$^2$tetrahedra}
\newlength{\myL}
\newcommand\rkhex[1]{%
  \setlength{\myL}{#1}
  \psset{unit=1.0}
  \begin{pspicture}(-0.9511\myL,-0.8090\myL)(0.9511\myL,\myL)
  \psline[linewidth=0.5pt](0.5\myL,-0.866\myL)(-0.5\myL,-0.866\myL)
  \psline[linewidth=0.5pt](-\myL,0)(-0.5\myL,0.866\myL)%
  \psline[linewidth=0.5pt](0.5\myL,0.866\myL)(\myL,0)%
  \end{pspicture}%
}
\newcommand\scalehex[1]{%
  \setlength{\myL}{#1}
  \psset{unit=1.0}
  \begin{pspicture}(-0.9511\myL,-0.8090\myL)(0.9511\myL,\myL)
  \psline[linewidth=0.5pt](0.5\myL,-0.866\myL)(-0.5\myL,-0.866\myL)(-\myL,0)(-0.5\myL,0.866\myL)%
  (0.5\myL,0.866\myL)(\myL,0)(0.5\myL,-0.866\myL)
  \end{pspicture}%
}
\newcommand\rkhextwo[1]{%
  \setlength{\myL}{#1}
  \psset{unit=1.0}
  \begin{pspicture}(-0.9511\myL,-0.8090\myL)(0.9511\myL,\myL)
  \psline[linewidth=0.5pt](-\myL,0)(-0.5\myL,0.866\myL)%
  \psline[linewidth=0.5pt](0.5\myL,0.866\myL)(\myL,0)%
  \end{pspicture}%
}

\newcommand\oep[1]{$O(\spa^#1)$} 

\newcommand{\umat}{{\sf 1\hspace*{-0.35ex}\rule{0.15ex}{1.6ex}\hspace*{0.35ex}}}

\newcommand{\nhex}{n_{\hexagon}}

\newcommand{\beq}{\begin{equation}}
\newcommand{\eeq}{\end{equation}}
\newcommand{\bea}{\begin{eqnarray}}
\newcommand{\eea}{\end{eqnarray}}

\newcommand{\etal}{{\em et al.}}

\newcommand{\spn}{{\mathrm{Sp}(N)}}
\newcommand{\bu}{b_\uparrow}
\newcommand{\bd}{b_\downarrow}
\newcommand{\da}{^\dagger}
\newcommand{\spa}{\varepsilon}

\def\state#1{\left|{#1}\right>}
\def\etats#1{\left<{#1}\right|}

\def\tit#1#2#3#4#5{{#1}{\bf #2}, #3 (#4)}

\def\prl{Phys.\ Rev.\ Lett.\ }

\def\prb{Phys.\ Rev.\ B\ }

\def\jap{J.\ Appl.\ Phys.\ }
\def\zpb{Z.\ Phys.\ B\ }
\def\jpsj{J.\ Phys.\ Soc.\ Jpn.\ }

\def\cjp{Can.\ J. Phys.\ }

\def\state#1{\left|{#1}\right>}

\def\bei{\begin{itemize}}
\def\eei{\end{itemize}}

\begin{document}


\title{Quantum dimer models and effective Hamiltonians 
on the pyrochlore lattice}

\author{R. Moessner,$^1$  
S. L. Sondhi$^2$ and M. O. Goerbig$^3$}

\affiliation{$^1$Laboratoire de Physique Th\'eorique de l'Ecole Normale
Sup\'erieure, CNRS-UMR8549, Paris}

\affiliation{$^2$Department of Physics, Princeton University,
Princeton, New Jersey}

\affiliation{$^3$Laboratoire de Physique Th\'eorique et Hautes Energies,
CNRS-UMR 7589, Universit\'e Paris 6 et 7, Paris}

\date{\today}

\begin{abstract}
We study a large-$N$ deformation of the $S=1/2$ pyrochlore Heisenberg
antiferromagnet which leads to a soluble quantum dimer model at
leading non-trivial order. In this limit, the ground state manifold --
while extensively degenerate -- breaks the inversion symmetry of the
lattice, which implies a finite temperature Ising transition
without translational symmetry breaking. At lower temperatures and
further in the $1/N$ expansion, we discuss an effective Hamiltonian
within the degenerate manifold, which has a transparent physical
interpretation as representing dimer {\em potential} energies. We find
mean-field ground states of the effective Hamiltonian which exhibit
translational symmetry breaking. The entire scenario offers a new
perspective on previous treatments of the SU(2) problem not controlled
by a small parameter,
in particular showing that a mean-field state considered previously
encodes the physics of a maximally flippable dimer configuration. We
also comment on the difficulties of extending our results to the SU(2)
case, and note implications for classical dimer models.
\end{abstract}


\maketitle

\section{Introduction}
One of the central open questions in the study of frustrated systems
is what happens when a classically highly degenerate magnet is
subjected to violent quantum fluctuations. A model system which has
played an important role in the discussion of this question
is the nearest neighbour
spin 1/2 Heisenberg antiferromagnet on the
pyrochlore lattice.
This lattice consists of corner-sharing tetrahedra and exhibits a
massive classical degeneracy.\cite{pyro-classic}

Very little is known reliably about the quantum 
model: exact solutions are
unavailable in $d=3+1$, Monte Carlo simulations are frustrated by the
sign problem, and the pyrochlore lattice -- being three dimensional
and having a unit cell of four spins -- does not yet 
lend itself to exact diagonalisations.

\begin{figure}
\epsfysize+3cm
\epsffile{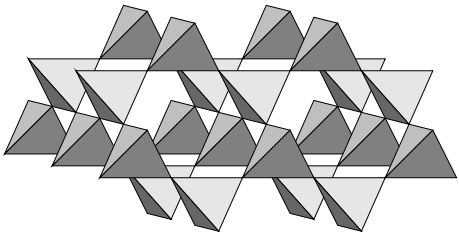}
\caption{The pyrochlore lattice, a network of corner-sharing tetrahedra.
The tetrahedra on the two (`up' and 'down') sublattices are shaded
differently.  The centres of the tetrahedra of either sublattice form
a face-centred cubic lattice.}
\label{fig:pyro3d}
\end{figure}

In an important, if somewhat cryptic, initial piece of analytical work
back in 1992,\cite{hbb} Harris, Berlinsky and Bruder (HBB) took the
approach of considering the bonds belonging to one of the two
sublattices -- ``weak bonds'' -- of tetrahedra perturbatively (the
tetrahedra of the pyrochlore lattice are arranged on the bipartite
diamond lattice, which in the current context can usefully be thought
of as two interpenetrating face-centred cubic lattices, see
Fig.~\ref{fig:pyro3d}). The
tetrahedra on the other sublattice, where the ``strong bonds'' reside,
are thus initially decoupled, with doubly degenerate dimerised singlet
ground states, parametrised 
by a pseudospin $\sigma$.  The basic idea of HBB
is then to switch on the weak bonds perturbatively, thus generating an
effective interaction between the $\sigma$s on different sites. This
interaction determines the eventual ground state, and in
itself defines a difficult quantum problem.

A number of authors have since developed ideas based on expansions for
pseudospins on an fcc lattice.\cite{tsune,canalstet,eea} Tsunetsugu's
detailed work extended the scope of the mean-field theory,\cite{tsune}
while Berg, Altman and Auerbach implemented a sophisticated numerical
procedure projecting the Hamiltonian onto this Hilbert
space.\cite{eea}

These
approaches have in common that their starting point has a lower
symmetry than the initial Hamiltonian, as a distinction between the
two sublattices of tetrahedra discards the inversion symmetry of the
pyrochlore lattice. In the same spirit, a perturbation theory in the
weak bonds but for the full density matrix, which unlike the previous
papers does not impose a restriction to the dimerised singlet
subspace, at any rate found spin ordering to be absent but did not
study the presence of bond ordering.\cite{canalslacroix}

In this paper, we take a different route by considering a Hamiltonian
with an enlarged internal, rather than reduced point, symmetry: we
study the spin-1/2 problem on the pyrochlore lattice via a large-$N$
quantum dimer model approach. This approach does not expressly break
the symmetry between the two sublattices of tetrahedra. It provides an
(artificial) small parameter, $1/N$, which we will use to obtain an
analytical solution to the first non-trivial order of the 
quantum dimer model.

Very unusually, we find that the quantum dimer model generated at
$O(1/N)$ is exactly soluble.\cite{fn-othersoluble} There turns out to
be a phase transition in the {\em Ising} univsersality class at a
temperature $T\sim O(1/N)$, where the inversion symmetry of the
pyrochlore lattice is spontaneously discarded. The set of states
selected at this order preserves an extensive entropy.

The question what happens within this subspace at higher order in
$1/N$ is closely analogous to that posed in
Refs.~\onlinecite{hbb,tsune,canalstet,eea}, as the respective starting
Hilbert spaces are isomorphic. Our approach thus provides a way of
reaching this starting point dynamically, i.e.\ by spontaneous rather
than explicit symmetry breaking.

We then show that the effective Hamiltonians obtained in
Refs.~\onlinecite{hbb,tsune,canalstet,eea} belong to a family which
has a simple interpretation as dimer potential terms in such an
approach; as the natural basis beyond $O(1/N)$ is not a dimer one,
however, they define a quantum Hamiltonian.

These Hamiltonians have been studied using different
approaches.\cite{hbb,tsune,canalstet,eea} We show that a
supertetrahedral ordering pattern proposed by HBB and Tsunetsugu can
be represented by a simple maximally flippable dimer configuration. We
also show that, within the framework of the mean-field theory, there
are further configurations close by in energy, leading to numerically
small characteristic energies. This implies that if there is
translational symmetry breaking, the final ordering pattern might not
be assumed until a temperature low compared to finite-size gaps in
exact diagonalisations or to other perturbations in real compounds.

Our results thus provide an intuitive picture of the dominant physics
captured in a class of theories proposed for the pyrochlore quantum
antiferromagnet.  As a byproduct, our results also imply that a
classical dimer model on the pyrochlore lattice with a simple
potential term will have an inversion and translation symmetry broken
ground state with residual entropy.

In the remainder of this paper, we first introduce the $\spn$ quantum
dimer model (Sect.~\ref{sec:spndim}), followed by its exact solution
(Sect.~\ref{sec:spnexact}). In Sect.~\ref{sec:spnhigher}, we discuss
the structure of the problem at higher orders in $1/N$ and in
Sect.~\ref{sec:spnmaxflip} make a connection to previous work via
the idea of maximally flippable dimer configurations. After some
remarks on further extensions (higher-order loops, other lattices and
SU(2) spins, Sects.~\ref{sec:spnlongloops}-\ref{sec:spnotherlatt}), we
close with a short discussion of further questions raised by this
approach.

\section{The $\spn$ dimer model}
\label{sec:spndim}

An $\spn$ dimer model is obtained from an SU(2) spin problem via
enlarging the symmetry of spin space from SU(2)$\sim$Sp(1) to
$\spn$. Rewriting the Heisenberg Hamiltonian in terms of operators
$s_{ij}^\dagger$, which create a singlet on the bond between spins $i$
and $j$, gives
\bea
H=-J \sum_{\left<ij\right>}
\left\{
s_{ij}\da s_{ij}-\frac1{4N^2}
\right\} ,
\eea
where $N=1$ for SU(2)$\sim$Sp(1).  The same form of $H$ holds for the
generalisation to $\spn$, with $s_{ij}^\dagger$ now creating $\spn$
singlets. Details for this and the following steps are provided in the
Appendix.

\subsection{The dimer model at leading order}

A hardcore dimer model is obtained from this Hamiltonian at leading
order\cite{subirnotes} as one takes
$N\rightarrow\infty$.\cite{fn:difference} At this order, any
nearest-neighbour dimer covering is an {\em orthogonal, degenerate
eigenfunction} of the $\spn$ Hamiltonian, with the ground state
energy per site of $E_0=-J/2$.

The ground state is thus exponentially degenerate in the volume of the
system, although the precise value of the degeneracy for three
dimensional lattices is not known. 
As the temperature is lowered below
$T\sim 1$, the $\spn$ magnet enters the dimer manifold of ground
states. It is possible in principle that this manifold already
incorporates some form of order, in which case this would happen via a
phase transition. However, it appears more likely that the
correlations averaged over the dimer manifold remain short-ranged for
the case of the non-bipartite pyrochlore lattice,\cite{hkms} 
in which case the
restriction to the dimer manifold has the form of a crossover.

The dimer wavefunctions obtained here are isotropic in spin
space. This means that any further symmetry breaking occurring at
subleading order can only be of a spatial variety. 

\subsection{Derivation of the quantum dimer model}
The degeneracy between different dimer coverings exists only at
leading order; at higher order, the model acquires a non-trivial
quantum dynamics. This dynamics is determined along the lines
pioneered by Rokhsar and Kivelson in the context of the SU(2)
Heisenberg model.\cite{Rokhsar88} They derived a Hamiltonian in the
space of dimer coverings by formally carrying out an expansion in a
parameter $x$ which in fact has the finite value of 1/2, as we will
briefly describe below.\cite{Rokhsar88,kivunpub} In the $\spn$ model,
this parameter is (artificially but) truly small:
$\spa\equiv1/2N$. The derivation of the dimer Hamiltonian is hence
completely analogous, albeit rigorously organised, in the orders of
the small parameter. Details are again given in the appendix.

At \oep{1}, the resulting quantum dimer Hamiltonian is very simple. It
reads:
\bea
H_{QDM}=2J\spa\Box\ .
\label{eq:Hmat}
\eea
Here, $\Box$ stands for a resonance term around a closed loop of
length four, i.e.\ a kinetic term which exchanges occupied and empty
links if they alternate around such a loop.  Note that this quantum
dimer model, unlike the Rokhsar-Kivelson one,\cite{Rokhsar88} only
contains a kinetic term.\cite{subirnotes}

\section{Exact solution of the QDM: partial
order by disorder} 
\label{sec:spnexact}
The quantum dynamics induced by Eq.~\ref{eq:Hmat}
can only lead to resonances between two dimer configurations which can
be transformed into one another by moving exactly two dimers. Such
moves are only possible for two dimers on a single tetrahedron, as the
shortest closed loop not confined to a single tetrahedron has length
six and therefore would require moving three dimers in order to
satisfy the hardcore constraint in both the initial and final
configurations. On a single tetrahedron, there are three possible
dimerisations in $\spn$, and the Hamiltonian matrix elements between
them read, at this order:
\bea
H_{tet}=2J\spa \left( \matrix{ 0 & 1 & 1 \cr 1 & 0 & 1 \cr 1 & 1 & 0 \cr  }
\right)\ .
\label{eq:tetres}
\eea
This matrix has one non-degenerate eigenvalue $4J\spa$ 
with corresponding
eigenvector $\state{\phi_A}=(1,1,1)/\sqrt{3}$. The other two
eigenvalues are degenerate at $-2J\spa$ and eigenvectors
$\state{\phi_+}=(1,\chi,\chi^2)/\sqrt{3}$ and
$\state{\phi_-}=(1,\chi^2,\chi)/\sqrt{3}$, with $\chi\equiv\exp(2\pi
i/3)$. These two eigenvectors form an $E$ representation of the
tetrahedral group $T_d$. Note that the asymmetry of the spectrum under
$J\leftrightarrow -J$ is a manifestation
of our inability -- in contrast to the 
case of the square\cite{Rokhsar88} 
and triangular\cite{MStrirvb} lattices -- to choose the
sign of the overlap matrix elements at will.

A single tetrahedron at this order is thus occupied by 0 or 1 dimers
and gains no energy from resonance moves, or it is occupied by 2
dimers, in which case it gains an energy $-2J\spa<0$.  We thus need to
maximise the number of tetrahedra occupied by two dimers. 

As the number of dimers equals the number of tetrahedra, this is done
by putting no dimers on one half the tetrahedra, and two each on the
remaining tetrahedra. As two tetrahedra with two dimers each cannot be
neighbours, and as the tetrahedra reside on the bipartite diamond
lattice, this implies that one sublattice of tetrahedra, containing
the `down' tetrahedra, say, is empty, whereas the `up' tetrahedra have
two dimers each. This gains an energy of $-2J\spa$ per up
tetrahedron, and hence $-J\spa/2$ per site.

The quantum dimer model up to this order thus spontaneously breaks the
inversion symmetry through the sites of the pyrochlore lattice (this
operation exchanges up and down tetrahedra); the point group is broken
down from $O_h$ to $T_d$. This is, to our knowledge, the first model
where such a symmetry breaking can be demonstrated to occur; this
feature is not unimportant as most treatments use this symmetry
breaking as part of the starting assumptions. We also note that this
is a rare case where a quantum dimer model is analytically soluble --
normally, finding the solution is a hard problem requiring intensive
numerics in itself.\cite{fn-othersoluble}

Despite this symmetry breaking, the ground state at \oep{1} retains a
large degeneracy as each single tetrahedron continues to exhibit a
two-fold degeneracy; the residual entropy per spin
is hence ${\cal S}=(k_B/4)
\ln2$, where $k_B$ is the Boltzmann constant.

We thus find a transition  at which the point group is
reduced by discarding inversion symmetry, but spin rotational as well
as real space translational symmetries remain intact. This is an
example of (only partial) order by disorder in a quantum frustrated
system.

\section{The fate of the residual degeneracy}
\label{sec:spnhigher}
So far, to $O(\spa)$, we have been fortunate in dealing with a quantum
dimer model with a simple structure. Configurations differing in the
distribution of dimers between tetrahedra define disconnected
dynamical sectors, the ground-state energies of which are given by the
number of doubly-occupied tetrahedra, with the corresponding
wavefunctions being outer products over tetrahedron
wavefunctions. Note that such a simple structure is generically absent
for other lattices, the leading dimer model typically not being
exactly soluble.

Here, these problems are deferred to the next order, $\spa^2$, where
the degrees of freedom are given by one pseudospins-1/2 for each
doubly occupied tetrahedron. For the ground-state sector, these
pseudospins reside on the face-centred cubic lattice defined by one
sublattice of tetrahedra. These pseudospins encode the
$E$-representation of $T_d$ provided by the degenerate wavefunctions
$\state{\phi_\pm}$.

Nonetheless, further insight can be gained by continuing to use a
dimer basis, on the understanding that a diagonal operator in the
dimer basis will not be diagonal in the pseudospin basis. This
approach will lead us to a simple interpretation of the effective
Hamiltonians of Refs.~\onlinecite{hbb,tsune,canalstet,eea}. Such a
connection is possible as, formally, the pseudospins on a
face-centred cubic lattice are also the starting point of those
studies, where the three (linearly dependent) SU(2) dimer coverings of
an isolated tetrahedron reduce to a two-dimensional
$E$-representation. 

\subsection{Effective Hamiltonians}
Let us consider the possible dimer operators we can write down beyond
the ones for the loops of length 4 already included at $O(\spa)$.  The
next simple kinetic term is that for a hexagon, $\hexagon$, which 
is in
principle generated at $O(\spa^2)$. However, such a term moves dimers
from the occupied sublattice of tetrahedra to the unoccupied one, and
thus does not have a matrix element of $O(1)$ within the pseudospin
subspace in which we are doing degenerate perturbation theory. It
will, however, contribute via a `virtual' process by first shifting
three dimers onto the empty tetrahedra and then back again. This
process is diagonal in dimer basis, pictorially represented by a
potential term of the form
\bea
H_{\hexagon}=-\sum_{\hexagon}\state{\rkhex{0.4em}}\etats{\rkhex{0.4em}} \ ,
\label{eq:hexpot}
\eea
where the symbol $\state{\rkhex{0.4em}}\etats{\rkhex{0.4em}}$ stands
for an operator which projects out the states with zero amplitude of
dimer occupancy on the three bonds of the occupied tetrahedra of a
given hexagonal loop. Here, the symbol $\sum_{\hexagon}$ implies a sum
over all such hexagonal loops.

The physics of this term is most transparent in the dimer basis.
However, it is off-diagonal in the basis of states we are doing
perturbation theory in. Let us therefore rewrite the dimer potential
term in pseudospin basis.

To do this, we write the wavefunction on each occupied tetrahedron
in terms of two angles, $\alpha$ and $\theta$:
\bea
\state{\theta,\alpha}&\equiv&
\sin(\alpha/2)\exp(i\theta/2)\state{\phi_+}\nonumber\\
&&+\cos(\alpha/2)\exp(-i\theta/2)\state{\phi_-}\ .
\eea
Next, we observe that the probability of finding a dimer on a given
link of the tetrahedron is given by
\bea
\left[1+\sin\alpha_i\cos(\theta_i-\nu_{\scalehex{0.06cm},i})\right]/3\ ,
\label{eq:sp2singlet}
\eea
where $\nu_{\scalehex{0.06cm},i}=0,\pm 2\pi/3$ according to which of
the three possible dimer pairings is selected.

Using this parametrisation, the potential term for the pseudospins on 
the face-centred cubic lattice has the form 
\bea
H_{fcc}=-\sum_{\hexagon}\prod_{i=1}^3
\left[1+\sin\alpha_i\cos(\theta_i-\nu_{\scalehex{0.06cm},i})\right]/3\ .
\label{eq:psinglet}
\eea
Here, the angles $\nu_{\scalehex{0.06cm},i}$ depend on the loop under
consideration and are
such that $\theta=\nu_{\scalehex{0.06cm},i}$ together with
$\alpha=\pi/2$ corresponds to the state which maximises the dimer
amplitude on the link which forms part of the hexagonal loop.
Put another way,
the factors in the product are the probabilities of finding 
dimers on the respective links. The sum on $i$ runs over the tetrahedra
to which these links belong.

This form can in turn be rewritten by introducing pseudospins
(represented by Pauli matrices $\sigma$) on each tetrahedron with the
quantisation axis along the azimuthal axis $\alpha=0$, and
$\hat{e}_{\scalehex{0.06cm},i}$ denoting the preferred directions
$\theta=\nu_{\scalehex{0.06cm},i}$ and $\alpha=\pi/2$: 
\bea
H_{fcc}=-\sum_{\hexagon}\prod_{i=1}^3
\left[(1+\Upsilon)+
\vec{\sigma}\cdot\hat{e}_{\scalehex{0.06cm},i}\right]/3\ .  
\label{eq:Hfcc}
\eea 
For the hexagonal potential, $H_{\hexagon}$, one has
$\Upsilon=0$. $\Upsilon\neq0$ corresponds to an additional potential
term involving only two dimers on a hexagon, pictorially represented
as $\state{\rkhextwo{0.4em}}\etats{\rkhextwo{0.4em}}\ $. 

The form of $H_{fcc}$ is in fact the same as that of the effective
Hamiltonians obtained in Refs.~\onlinecite{hbb,tsune,canalstet,eea}.
The intermediate steps in the algebra differ between Sp($N$) and SU(2)
on account of the non-orthogonality of the dimerisations of a
tetrahedron in SU(2) [see Sect.~\ref{sect:su2limit}].  Most
importantly, the $\nu_{\scalehex{0.06cm},i}$ are shifted by
$\pi$. 

However, in the end the effective SU(2) Hamiltonians correspond to
dimer potentials like $H_{\hexagon}$ (Eq.~\ref{eq:hexpot}) in the same
way that $H_{fcc}$ does. This common equivalence states in a crisp
geometrical form that both the Sp($N$) (Eq.~\ref{eq:Hfcc}) and the
SU(2) effective Hamiltonians of
Refs.~\onlinecite{hbb,tsune,canalstet,eea} encode dimer potentials.

\subsection{Mean-field theory for $H_{fcc}$}
Even with the restriction 
to this class of Hamiltonians, one is still
left with the hard problem of minimising $H_{fcc}$, which has the form of
a $S=1/2$ Heisenberg model with spin-orbit coupling.

A proper quantum mechanical treatment of this problem is beyond the
scope of this paper.  Rather, to shed some light on the physics
unearthed in previous studies, we considered this problem in a
mean-field theory, treating the pseudospins-1/2 as being effectively
classical.\cite{hbb,tsune,canalstet}

For $\Upsilon=0$, the best state we have discovered so far has a four
sublattice structure, as do the states of
Refs.~\onlinecite{hbb,tsune,canalstet,eea}, with angles
$\alpha\equiv\pi/2$ and $\{\theta_i\}=\{0,0,\pi-\delta,\pi+\delta\}$,
where $\delta=\arccos(7/8)\approx0.5054$.

This state has an energy per loop (and therefore per spin), $\lambda$,
with $\lambda=-289/3456\approx-0.08362$.  The highest possible energy,
realised for $\theta_i\equiv\pi$, gives $\lambda=0$. Even a state with
$\delta$ significantly different can have a closeby energy: for
example, for $\delta=0$, one obtains
$\lambda=-1/12\approx-0.08333$. This state has a higher symmetry as it
only represents a two-sublattice ordering.

As $\Upsilon$ is changed, $\delta$ evolves, and it reaches
$\delta=\pi/3$ at $\Upsilon=-1/2$; at this point, it in fact becomes
possible to disorder one of the four sublattices at zero cost in
energy; this we call Harris-Bruder-Berlinsky--Tsunetsugu (HBB-T)
state. 
For
$\Upsilon=0$, this configuration
$\{\theta_i\}=\{0,\theta_2,2\pi/3,-2\pi/3\}$ is no longer optimal in
our case but still rather close: $\lambda=-35/432\approx-0.08102$.

These results raise two points. The first is the observation that the
particular four-sublattice ordering with one disordered sublattice is
not entirely generic, as even in mean-field theory, it does not
correspond to an extended parameter region in $\Upsilon$.

The second is the recurring theme of the emergence of a numerically
small energy scale. There are configurations the symmetry properties
of which differ significantly but whose energies are very similar. The
configurations discussed above are degenerate at \oep{0} and
\oep{1}, and their variational energies with respect to  $H_{fcc}$ 
differ only by a few percent.

These facts suggest that the magnet will break translational symmetry
-- by assuming whatever ordered configuration it chooses -- only at
very low temperatures, provided quantum fluctuations do not prevent
this ordering altogether.

\section{The HBB-T state as a dimer state}
\label{sec:spnmaxflip}
Given the Hamiltonian $H_{fcc}$ is most simply written in dimer basis,
it is natural to ask whether the mean-field ground states have a
natural interpretation in terms of dimer coverings. This is most
pertinent for states with angles $\alpha=\pi/2$ and $\theta$ integer
multiples $2\pi/3$, as these correspond to maximal dimer amplitudes on
one of the three pairs of links,\cite{tsune} and are thus naturally
identified with a dimer configuration.

The HBB-T state in fact has a very simple such interpretation: it can
be represented as the dimer configuration with the maximal number of
flippable hexagons, that is to say as the ground state of the
classical dimer model with a hexagonal potential of the type
$H_{\hexagon}$ [Eq. (\ref{eq:hexpot})]. Why one sublattice of
tetrahedra remains disordered for the maximally flippable
configuration also becomes apparent in this language.

As an aside, we note that the appearance of such maximally flippable
configurations is a generic feature of quantum frustrated systems.  In
a nutshell, a large number of flippable loops implies a wide range of
possible fluctuations, and hence undetermined degrees of
freedom. These the perturbation can make use of by arranging them to
its liking. Such a route to quantum order by
disorder\cite{shenderobdo} has been discussed in detail in the context
of frustrated transverse field Ising models.\cite{kagtrfield}

To demonstrate this, we first need to show which classical dimer
states optimise this problem; then, we render the HBB-T state in a
way which makes its identity manifest.

\subsection{Maximally flippable configurations}
To maximise the number of flippable hexagons, $\nhex^{fl}$, we note
that the hexagons lie in the [111] kagome planes, and that each bond
of the pyrochlore lattice is part of hexagons in two different planes,
e.g. [111] and [--1--11]. Thus, each dimer can be part of at most two
flippable hexagons; as the number of hexagons, $\nhex$, equals the
number of spins, and hence twice the number of dimers, and as three
dimers are needed for a flippable hexagon, this provides an upper
limit of $\nhex^{fl}/\nhex=1/3$. 

\begin{figure}
\epsfysize+5.5cm
\epsffile{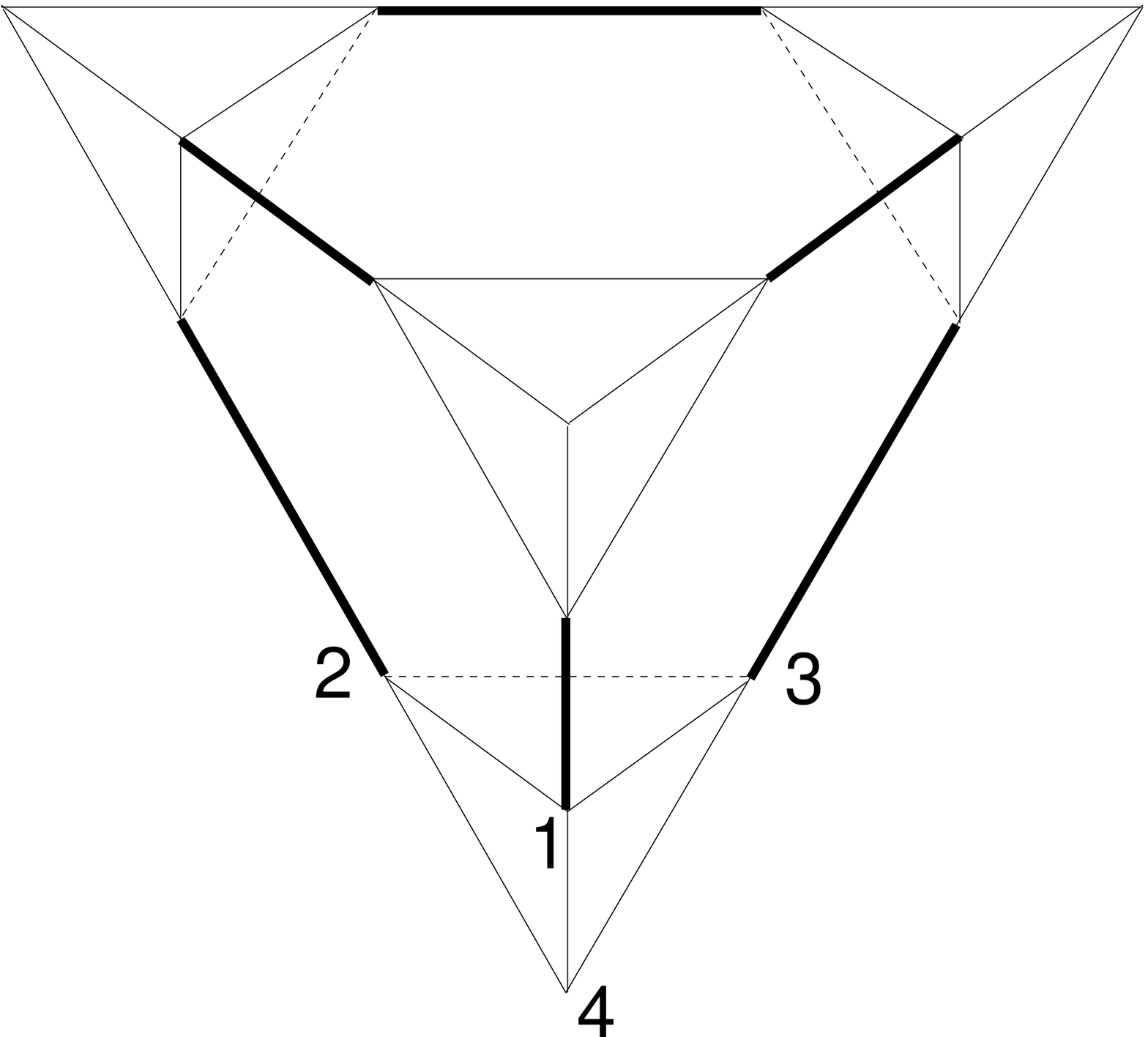}
\caption{A \ashtray~consists of six dimers, which form four flippable 
hexagons in the four $\{111\}$ 
kagome planes, and four surrounding 
dimerfree tetrahedra of the down-sublattice, such as 
(1234).}
\label{ashtray1}
\end{figure}

However, it turns out to be impossible to saturate this bound, as can
be seen by explicitly constructing configurations which are locally
maximally flippable.  Consider first a configuration with independently 
flippable loops. In 
this case, one needs at least three dimers per resonant loop, and one obtains
for the upper bound $\nhex^{fl}/\nhex=1/6$.
A more favourable state would be one in which two flippable hexagons share
one dimer in two different kagome planes, e.g. [111] and [--111]. For such
a local configuration one needs at least five dimers for each pair of 
resonant loops, and this yields the upper bound $\nhex^{fl}/\nhex=1/5$.
 
This local configuration might in principle be used as a building block
for a new configuration if other dimers are added locally to form other 
flippable hexagons. 
There is only one possibility of adding a sixth dimer to build a third
flippable hexagon, and this automatically yields a forth one, with
one flippable hexagon in each kagome plane. This local configuration
of four flippable hexagons, constructed by six dimers, has the form of
a \ashtray~shown in Fig.~\ref{ashtray1}.

This \ashtray~has four dimerfree tetrahedra at its corners.  None of
the dimers emanating from the outer vertices of these tetrahedra can
be part of a flippable hexagon, and thus at least 2 dimers are lost
(four dimers each shared between two \ashtrays). 

To obtain four flippable hexagons, one thus needs to invest $6+2=8$
dimers, and hence 16 sites. As the number of sites equals the number
of hexagonal loops, this establishes the upper bound of
$$\nhex^{fl}/\nhex \leq 1/4\ .$$

A consequence of the local \ashtray~structure is that an optimal
configuration, with $\nhex^{fl}/\nhex = 1/4$ automatically breaks the
inversion symmetry, i.e. all up-tetrahedra contain two dimers and all
down-tetrahedra are dimerfree. 

As an aside, we note that a classical attractive potential for dimer
pairs on a tetrahedron would have selected the dimer manifold (with
entropy $k_B \ln 3/4$ per site) with all dimers on one sublattice of
tetrahedra. The attractive hexagonal loop term (by itself or in
combination with this term) in addition imposes the creation of the
supertetrahedra.

\begin{figure}
\epsfysize+7.8cm
\epsffile{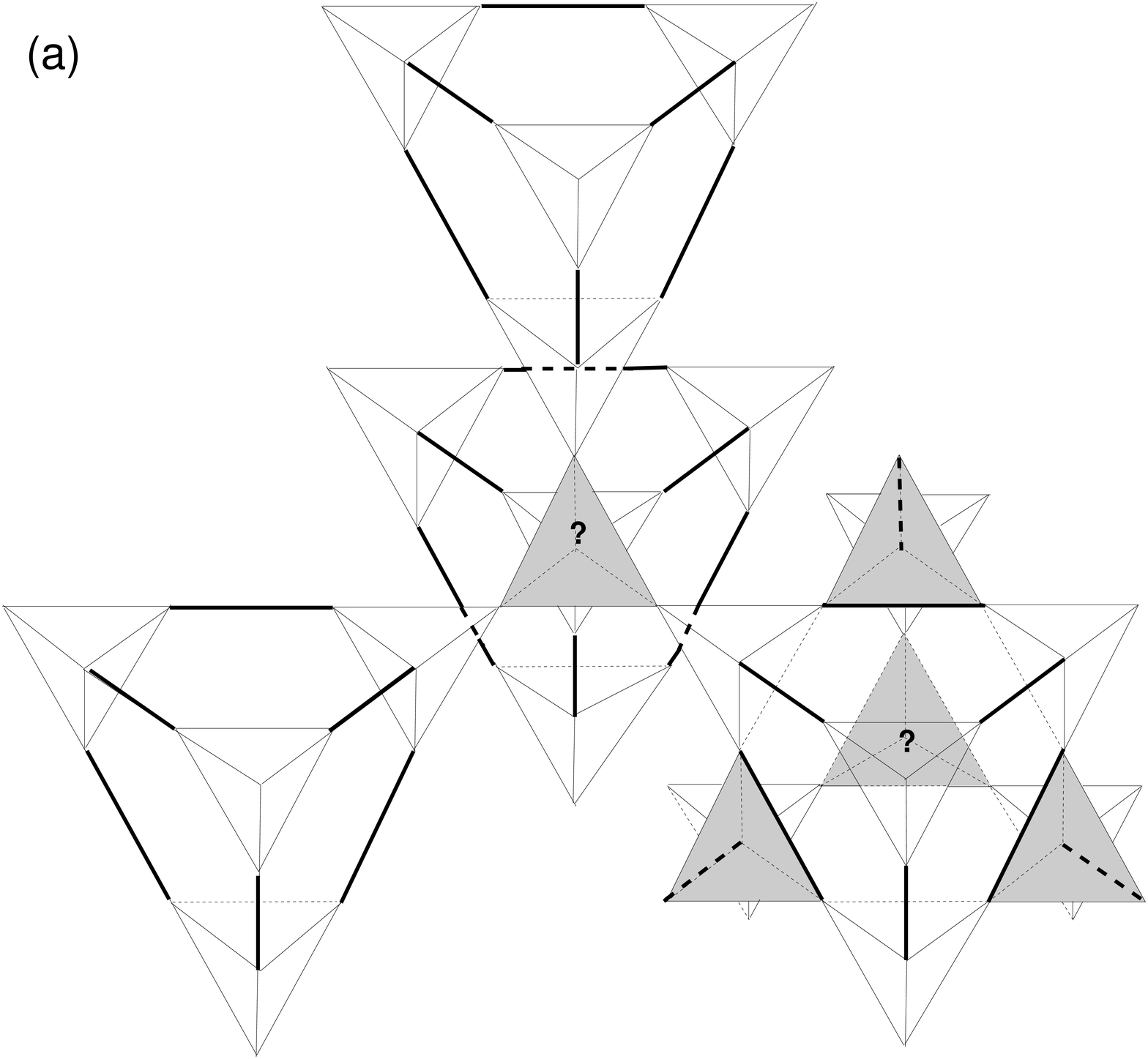}

\vspace*{0.3cm}
\epsfysize+6.5cm
\epsffile{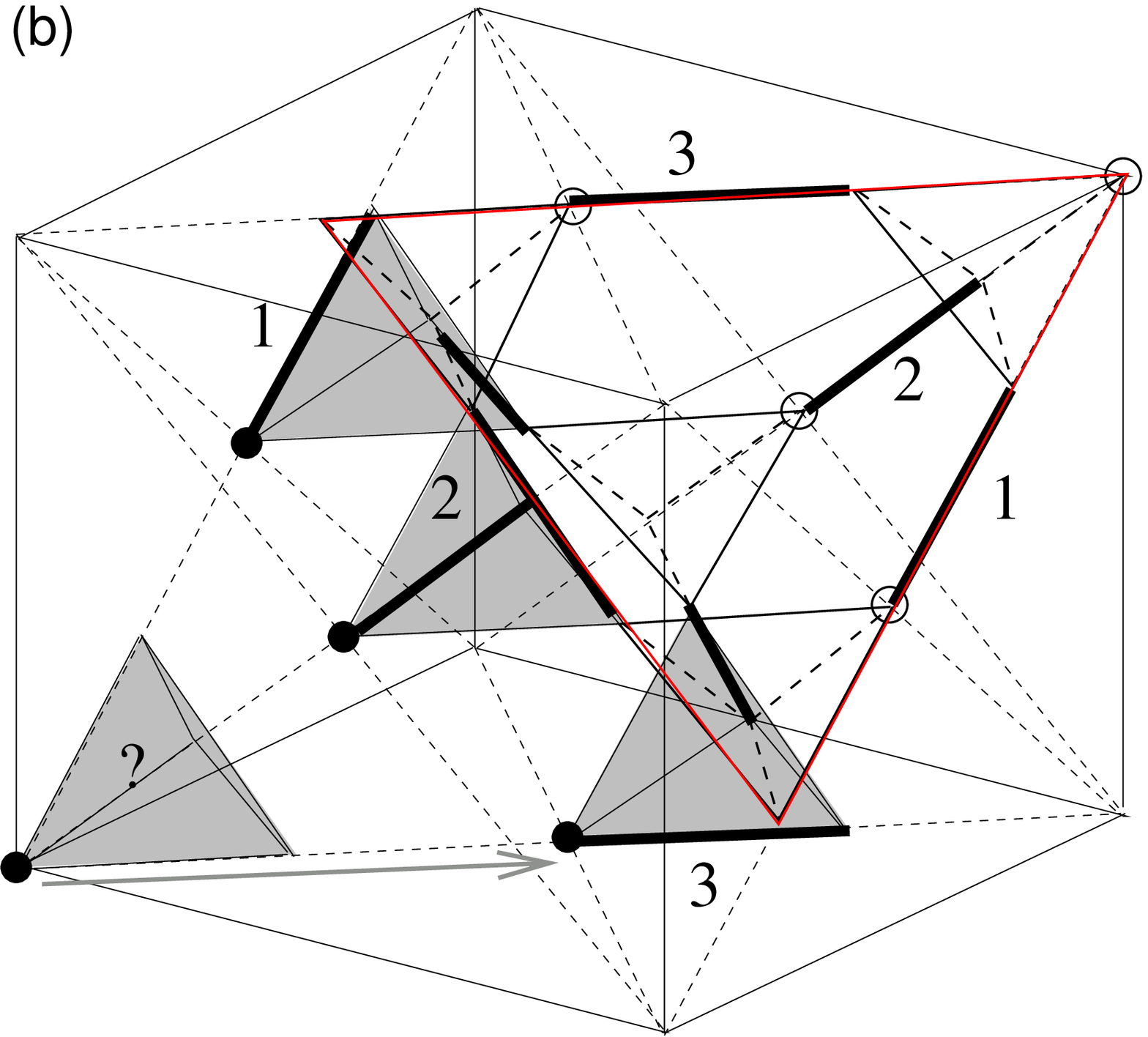}
\caption{Optimal dimer configuration. {\sl (a)} Gray tetrahedra form the 
up-sublattice, and white tetrahedra (dimer-free) form the
down-sublattice. Four \ashtrays~meet in
one ?-tetrahedron of the up-sublattice. The dimer configurations of
these ?-tetrahedra may be chosen arbitrarily. The \ashtray\ on the right
hand side is represented in such a manner to make clear the relation
to the mean-field configuration of Ref.~\onlinecite{tsune}. {\sl (b)} Cubic 
lattice cell of the \ashtray~configuration. The \ashtray~is shown in red.  }
\label{ashtray2}
\end{figure} 

\subsection{Connection to the HBB-T state}

An optimal \ashtray\ tiling is shown in Fig.~\ref{ashtray2}.  The
dimers which do not take part in the \ashtray~formation are found on
the ?-tetrahedra in Fig.~\ref{ashtray2}a and may be chosen freely.
The ground-state degeneracy of the optimal configuration therefore
remains macroscopic, and the entropy per site is
$\mathcal{S}_{\varepsilon^2}=(k_B/16)\ln 2$.\cite{tsune} Because the
up-tetrahedra form an fcc-lattice, the ?-tetrahedra now live on a
cubic lattice and connect the \ashtrays.  This is reminiscent of the
HBB-T state; indeed, the cartoon for this state is precisely our state
of \ashtrays, see Fig.~\ref{ashtray2}, where we have shown the
\ashtrays, together with the dimer configurations on the shaded
tetrahedra as dictated by the HBB-T state. However, by focussing on
the up-tetrahedra, rather than the flippable loops, the geometry is
harder to visualise.

One may ask whether there exist other tilings of \ashtrays which are
also optimal. Indeed, if one chooses the ?-tetrahedron in the cubic
lattice cell in Fig. \ref{ashtray2}b in such a manner that its
dimerisation coincides with that of the tetrahedron at lattice point
[110] (3-dimerisation), one may displace one layer of cubes with
respect to a neighbouring layer in the [110] direction, as indicated
by the gray arrow. 

However, this one choice forces us to fix the dimerisation of all
?-tetrahedra at the interface of the displacement. The resulting
interface energy will thus generically win over the displacement
entropy and lift this degeneracy.

\section{Higher-order loops}
\label{sec:spnlongloops}
In the absence of an exact solution at $O(\spa^2)$, our present
approach does not {\em per se} justify pursuing the physics of longer
loops. For completeness, however, we mention some geometric facts
about the entropics of the residual degrees of freedom in the dimer
model with the uniform four-sublattice ordered state, with one
disordered sublattice, as a starting point. We will discuss this topic
more completely separately.\cite{fn-mog}

\begin{figure}
\epsfysize+6.0cm
\epsffile{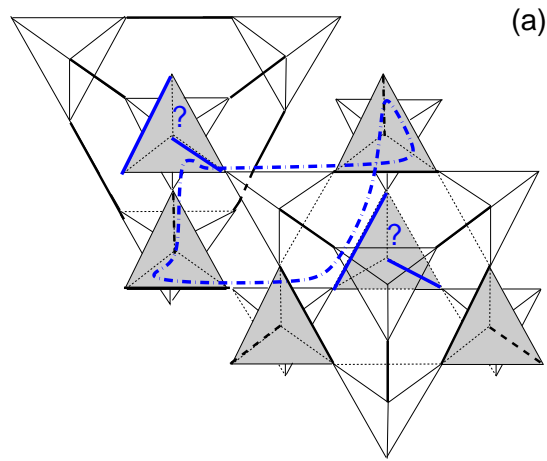}
\caption{Twelve-bond loop connecting two
nearest neighbours of ?-tetrahedra (blue dashed-dotted line). The
dimerisation of the connected ?-tetrahedra has to be the same ({\sl
here:} blue).}
\label{fig:longloopNN}
\end{figure}

\begin{figure}
\epsfysize+8.0cm
\epsffile{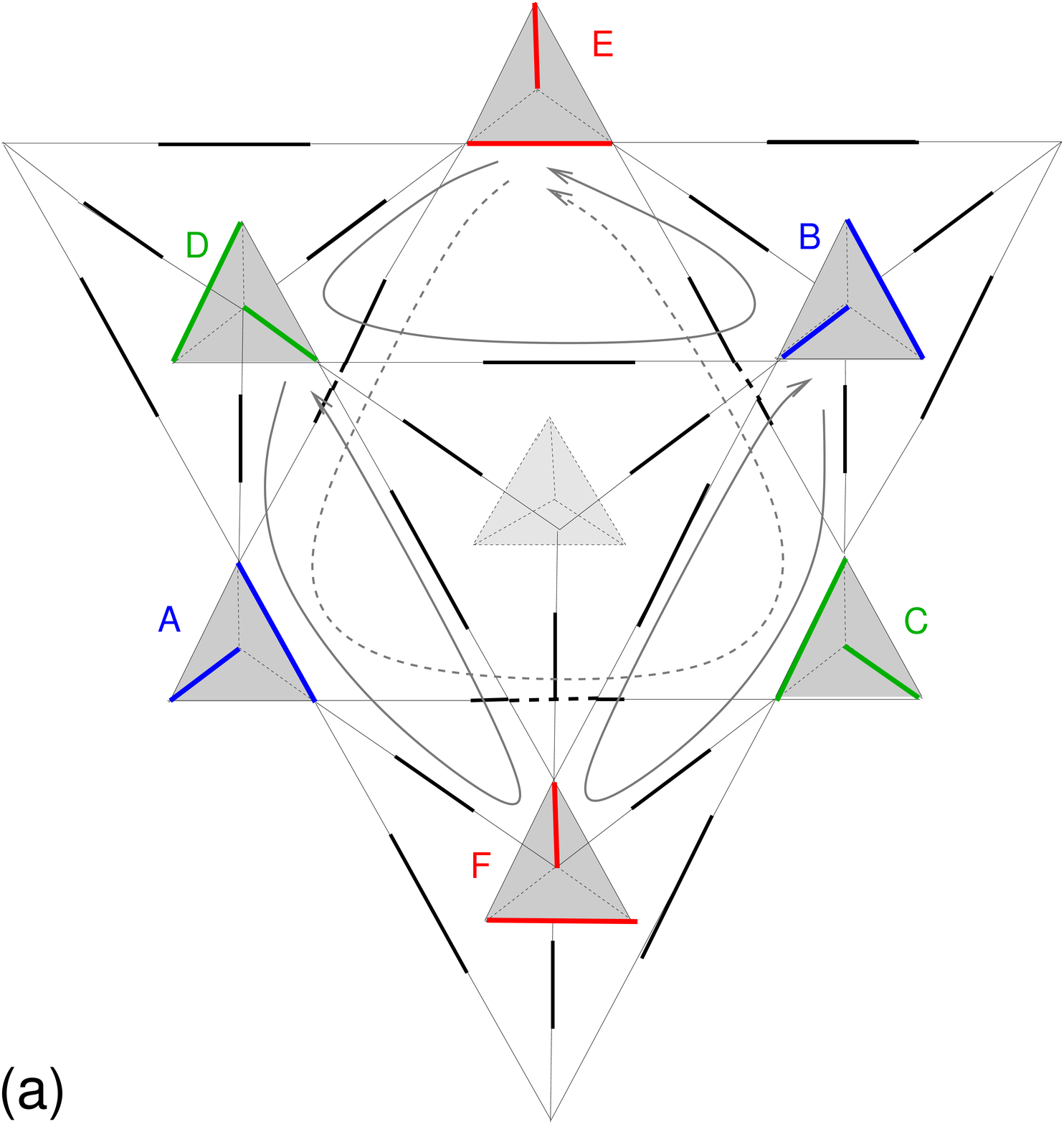}
\vspace*{0.3cm}
\epsfysize+5.5cm
\epsffile{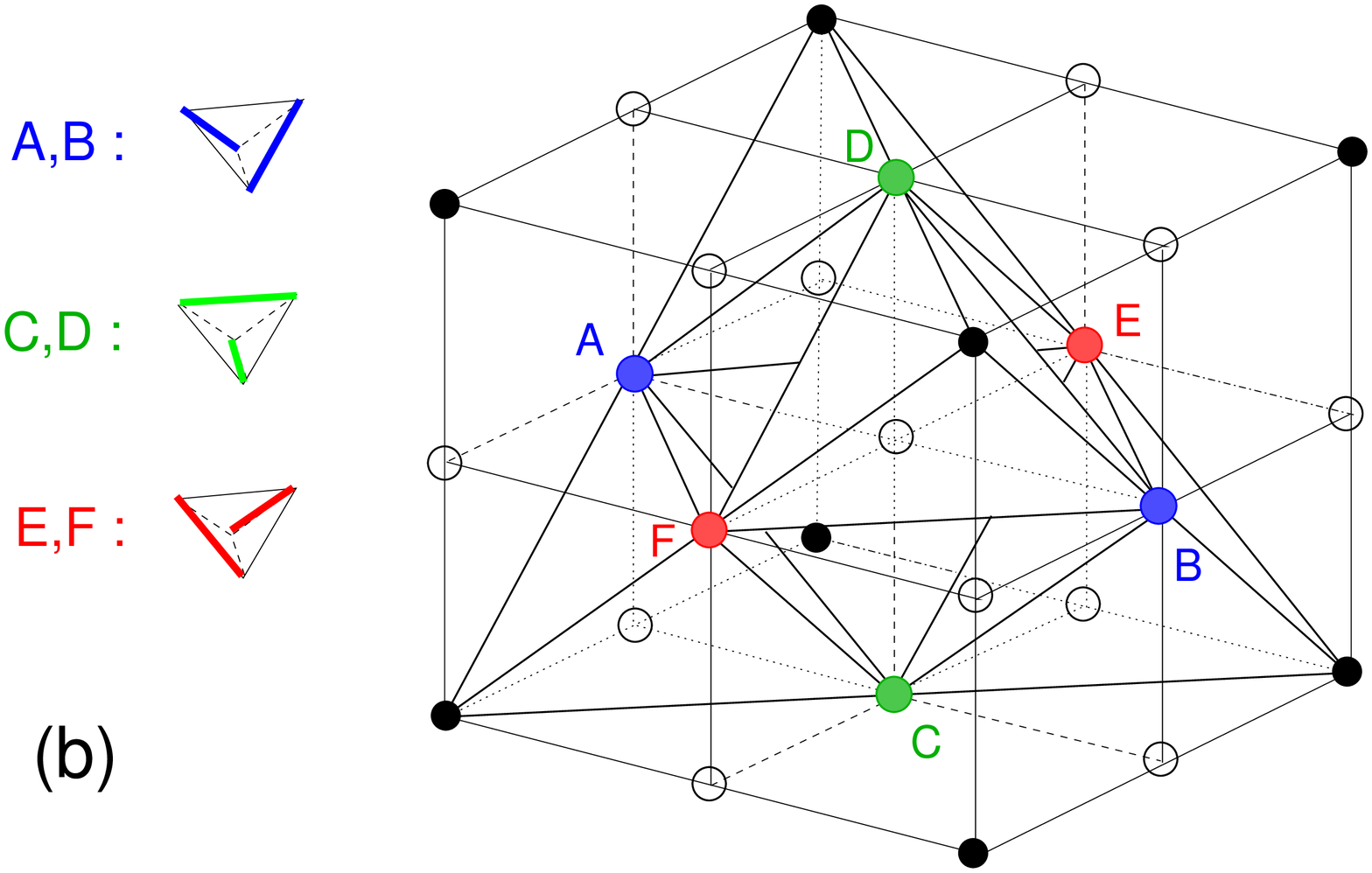}
\caption{Next-nearest-neighbour loops. (a) Closest packing of 12-bond loops 
connecting next-nearest neighbours of ?-tetrahedra (dark gray).  The
tetrahedron in light gray is part of the other square sublattice. The
superstructure is again an \ashtray, but with unequal bonds
(\superashtray).  (b) Elementary lattice cell of the
\superashtray~configuration. The ?-tetrahedra on the cubic sublattice,
which are connected by next-nearest-neighbour loops, are represented
by full circles (black and coloured). The open circles represent
?-tetrahedra on the second cubic sublattice. The four resonant
next-nearest-neighbour loops are those connecting $AFD$, $BED$, $BFC$,
and $AEC$. 
}
\label{fig:longloopnnn}
\end{figure}

The as yet undetermined dimers can only participate in virtual
resonance loops of length 12 (involving moving six dimers).  One,
rather contorted loop, can connect a pair of nearest-neighbour
?-tetrahedra (Fig.~\ref{fig:longloopNN}). The other involves three
?-tetrahedra which are mutually next-nearest neighbours. This is
displayed in Fig.~\ref{fig:longloopnnn}a.

The nearest-neighbour loop can only (virtually) resonate if the dimers
are oriented perpendicular to the line joining the ?-tetrahedra, and
it thus favours the formation of rods of ?-tetrahedra with identical
alignment, somewhat reminiscent of the supertetrahedral ordering
pattern found in Ref.~\onlinecite{eea}.

The next-nearest neighbour loops only connect tetrahedra living on the
same sublattice of the (bipartite) cubic lattice on which the
?-tetrahedra reside. Trying to maximise the number of flippable
next-nearest neighbour loops thus leads to two identical copies of the
same problem. Amusingly, the maximally flippable n.n.n.-loop
configuration is self-similar: one now finds \superashtrays, with one
out of four ?-tetrahedra remaining again undetermined
(Fig.~\ref{fig:longloopnnn}b).

The maximal total number of flippable loops of length 12 is obtained
by taking the two copies of this \superashtray\ state and arranging
the ?-tetrahedra to satisfy the rod-arrangement dictated by the
nearest neighbour loops.

\section{The SU(2) limit}
\label{sect:su2limit}
The Rokhsar-Kivelson quantum dimer model for SU(2) spins is based on
an expansion in the same small parameter, $\spa$, appearing in the
overlap matrix. However, this parameter is not in fact infinitesimal
-- 1/2, in fact -- and this leads to a certain arbitrariness in
assigning an order to a given matrix element as a change in its order
can be offset by a numerical factor 2. This is most plainly seen for
the case of the constant $1/4N^2\sim$ \oep{2} in the Hamiltonian,
which in the case of SU(2) simply becomes a constant $1/4\sim$
\oep{0}.  Here, we evaluate the matrix
elements between two dimer configurations exactly and assign them the
order of the overlap between the two configurations.

On the pyrochlore lattice, the main obstacle lies in the fact that
SU(2) dimerisations of a tetrahedron are not linearly independent. In
the quantum dimer model, this shows up in the form of a non-invertible
overlap matrix $S$. In such a situation, Eq.~\ref{eq:ortho} is no
longer meaningful.

One can adopt two different stances with respect to this
observation. Firstly, one can hope that -- by continuity -- the
large-$N$ physics will nonetheless be relevant to the
SU(2)$\sim$Sp($N=1$) case. Alternatively, one can decide that SU(2) is
special, and that the physics discussed so far in this paper will
qualitatively have to be modified for the $N=1$ case.

Here, it is useful to note that one also finds that the resonance
energy of a single tetrahedron vanishes -- for the SU(2) RK model, the
matrix elements in Eq.~\ref{eq:tetres} are zero.
Let us thus sidestep the orthogonality 
issue by ignoring the length four loops
altogether, with a leading term of the QDM proportional to the
hexagonal resonance term $-J\spa^2\hexagon$.  This is fundamentally
different from $H_{fcc}$ in that it represents a kinetic, and not a
potential, dimer term. The ground state of this quantum dimer model
(and whether it breaks the lattice inversion, or indeed any other,
symmetry) is not known.  Candidate variational states for this problem
can be constructed along the lines explored in
Ref.~\onlinecite{kagtrfield}. One such candidate is the maximally
\ashtray-configuration discussed above.

As the flippable hexagons there are not independently flippable, other
candidate states can be constructed to maximise the number of
maximally flippable hexagons; one example can be obtained from the
proposal of Ref.~\onlinecite{protectorate}.  This has one in six
hexagons flippable and does not break the inversion symmetry. 
We leave a detailed analysis of this quantum
dimer model as a subject for future study.

\section{Other lattices}
\label{sec:spnotherlatt}
The approach developed here can easily be extended to other
lattices. However, the exact solubility discovered here is in general
not encountered elsewhere. In fact, the situation can differ from the
pyrochlore one in several respects.\cite{fn-mog}

Firstly, the lattice under
consideration may not admit any dimer coverings. Secondly, all dimer
states may have a fixed number of flippable shortest loops of even
length, thereby rendering the leading order dimer model
trivial. Thirdly, the shortest loops may be inequivalent or
overlapping in a way so as to destroy the solubility. In addition, the
appropriate low-energy sector may at any rate be better described by a
different type of effective model.\cite{milamamb}

\section{Conclusion}
In conclusion, we have presented an exactly soluble deformation of the
highly frustrated quantum pyrochlore antiferromagnet. To our
knowledge, this provides the first instance in which the inversion
symmetry of the pyrochlore lattice by itself is spontaneously broken.

From an experimental perspective, this is perhaps the most interesting
observation in this paper: upon lowering the temperature from the
paramagnetic phase, the current scenario implies the presence of an
initial Ising transition into a non-magnetic phase with finite
entropy and full translational and rotational symmetries.

Of course, the presence of such a symmetry broken phase is in a sense
already implicit in the starting points of
Refs.~\onlinecite{hbb,tsune,canalstet,eea}s With respect to these, our
approach has provided a simple physical picture by shedding some light
on the `natural' degrees of freedom arising in such treatments, the
hexagonal resonating loops. This reinforces the idea of an non-magnetic 
ordering pattern with an enlarged unit cell.

Our work therefore goes some way towards providing a rationale for the
sequences of symmetry-breakings discussed
before.\cite{hbb,tsune,canalstet,eea} Whether or not the scenario
discussed here provides the appropriate framework for understanding
the nearest neighbour S=1/2 pyrochlore Heisenberg antiferromagnet
remains an open question.

\section*{Acknowledgements}
We would like to thank Ehud Altman, Assa Auerbach, Benoit Dou\c{c}ot,
Subir Sachdev and Hirokazu Tsunetsugu for useful conversations.  We
are very grateful to Oleg Tchernyshyov for those and for collaboration
on closely related work, and to Steve Kivelson for sharing his
unpublished notes. RM is grateful to the Aspen Center for Physics and
the Lorentz Center of Leiden University, where part of this work was
carried out. He is supported in part by by the Minist\`ere de la
Recherche et des Nouvelles Technologies with an ACI grant. MOG is
supported by the Swiss National Foundation for Scientific Research
under grant No. PBFR-106672. SLS would like to acknowledge support by
the NSF (DMR-9978074 and 0213706) and the David and Lucile Packard
Foundation.

\appendix
\section{The $\spn$ dimer model}
As a first
step, one introduces bosonic operators (``Schwinger bosons'')
$\left\{\bu,
\bd\right\}$, to represent the spin operators in the following way:
${S}^z=(\bu\da\bu-\bd\da\bd)/2$, ${S}^+=\bu\da\bd$. In order to
represent SU(2) spins 1/2, one needs to supplement the bosonic
description with the constraint
$
n_b\equiv\bu\da\bu+\bd\da\bd=1 \label{eq:constr}
$ on each site.

In terms of these operators, the antiferromagnetic nearest-neighbour
Heisenberg Hamiltonian becomes ($J>0$):
\bea
{H}&=&J\sum_{\left<ij\right>}{\mathbf{S}_i\cdot \mathbf{S}_j}\\
&=&\nonumber
-J\sum_{\left<ij\right>}
\left\{
(\epsilon^{\sigma\tau}b\da_{i\sigma}b\da_{j\tau}/\sqrt{2})
(\epsilon^{\mu\nu}b_{i\mu}b_{j\nu}/\sqrt{2})
-1/4 
\right\}\ ,
\eea 
where $\mathbf{S}_i$ are spin
1/2 SU(2) operators. This
expression is formally generalised to Sp($N$) by the introduction of
an additional flavour index for the bosons, which we label by a
capital letter. In $\spn$, there are $N$ such flavours; the case of
SU(2) corresponds to one flavour: SU(2)$\sim$Sp(1).  The $\spn$
generalisation, ${\cal J}^{\mu\nu}_{AB}$, of the Levi-Civita symbol
continues to be off-diagonal and antisymmetric in the spin index but
simply diagonal in the flavour index: ${\cal
J}^{\mu\nu}_{AB}=\epsilon^{\mu\nu}\delta_{AB}$. 
We define the singlet
(hereafter called dimer) annihilation operator for bond
$\left<ij\right>$: 
\bea
s_{ij}\equiv {\cal J}^{\sigma\tau}_{AB}
b_{i\sigma}^Ab_{j\tau}^B/\sqrt{2N}\ , 
\eea
where summation over spin and
colour indices is implicit.
The $\spn$ 
Hamiltonian reads
\bea
H=-J \sum_{\left<ij\right>}
\left\{
s_{ij}\da s_{ij}-\frac1{4N^2}
\right\} .
\eea


\subsection{The dimer model at leading order}

A hardcore dimer model is obtained from this Hamiltonian at leading
order\cite{subirnotes} as one takes
$N\rightarrow\infty$.\cite{fn:difference} To see this, consider the
properties of dimer coverings of the lattice. Let $P$ denote an
ordered pairing of the sites of the lattice and $\state{\psi_P}$ the
corresponding singlet wavefunction: $\state{\psi_P}\equiv\prod
s_{p_{2n-1}p_{2n}}\da\state{0}$, where $\state{0}$ is the state with
no bosons present and $p_{2n-1}$\ and $p_{2n}$ are the two members of
the $n^{{\rm th}}$ pair contained in $P$. $n$ runs from 1 to $N_s/2$,
where $N_s$ denotes the number of sites.

Any such state $\state{\psi_P}$ satisfies the constraint $n_b=1$ for
every site. At leading order, $O(1/N^0)$, these coverings are {\em
orthogonal, degenerate eigenfunctions} of the $\spn$ Hamiltonian:
$H\state{\psi_P}=E_0\state{\psi_P}$, with the ground state energy 
$E_0=-J N_s/2$. 
Due to the constraint on
the boson number, it is not possible to have more than one dimer per
site, and due to the form of the Hamiltonian, it is disadvantageous to
have less. At $O(1/N^0)$, the ground states are thus all the hardcore
dimer coverings, denoted by $P$, of the lattice.  

\subsection{Derivation of the quantum dimer model}
As the different dimer coverings are not exactly orthogonal, one
obtains an overlap matrix, $S$, between different coverings which has
the following schematic form:
\bea 
S=\umat-\spa\Box+\spa^2\hexagon+\spa^2\Box\times\Box + O(\spa^3)\ .  
\eea 
Here,
$\umat$ is the unit matrix, and the symbols $\Box, \hexagon$ denote a
nonvanishing matrix element between two dimer configurations differing
in the positions of two and three dimers, respectively.

The {\em signs} of the overlap matrix elements cannot be chosen
freely; in fact, there is no convention which uniformly yields a
positive sign in front of the $\Box$ term. This is in contrast to the 
case
of the square and triangular lattices, 
where this choice is a matter of convention.\cite{Rokhsar88,MStrirvb}
However, one is free to
choose the sign of the $\hexagon$ terms to be, for example, all
positive or all negative. 

We can now formally orthonormalise the
basis set by introducing basis states
$\state{\Psi_p}\equiv\sum_{p^\prime}
\left(S^{-1/2}\right)_{pp^\prime}\state{\psi_{p^\prime}}$. Here, 
$S^{-1/2}$ is the (symmetric) inverse square root of $S$, which can be
obtained from $S$ by a Taylor expansion in $\spa$. The
$\state{\Psi_p}$ can be labelled uniquely by a parent dimer
configuration $P$, as the other dimer states are admixed only to
higher order in $\spa$. Explicitly, orthonormality follows from
\bea
\left<\Psi_p\right|\left.\Psi_q\right>&=&\sum_{p^\prime
q^\prime}
S^{-1/2}_{pp^\prime}S^{-1/2}_{qq^\prime}
\left<\psi_{p^\prime}\right|\left.\psi_{q^\prime}\right>\nonumber\\ 
&=&
(S^{-1/2}SS^{-1/2})_{pq}=\umat_{pq}\ .
\label{eq:ortho}
\eea
 
The matrix elements of $H$ in the orthogonalised basis now read
\bea
\nonumber
{\cal H}_{pq}&\equiv&\left<\Psi_p\right|H\left|\Psi_q\right>=
\sum_{p^\prime q^\prime}S^{-1/2}_{pp^\prime}S^{-1/2}_{qq^\prime}
\left<\psi_{p^\prime}\right|H\left|\psi_{q^\prime}\right>
\\ &=&
(S^{-1/2}HS^{-1/2})_{pq}\ . 
\eea

This expression is useful as it is also possible to expand
$\left<\psi_{p^\prime}\right|H\left|\psi_{q^\prime}\right>$ in powers
of $\spa$. It turns out to be convenient to subtract off the energy
of a dimerised state, $-J(1-1/4N^2)N_s/2$, so that diagonal 
terms are absent:
$\left<\psi_{p^\prime}\right|H\left|\psi_{p^\prime}\right>=O(\spa^2)$, 
and the
expansion of the Hamiltonian matrix contains no terms of
\oep{0}. Thence,
\bea
H_{QDM}=S^{-1/2}HS^{-1/2}=2J\spa\Box+O(\spa^2).
\eea

\end{document}